\documentclass[usenatbib,useAMS]{mnras}

\usepackage{fancyhdr}
\usepackage{amsfonts}
\usepackage{amsmath}
\usepackage{amssymb}
\usepackage{multicol}
\usepackage{layout}
\usepackage{graphicx}
\usepackage{epsfig}
\usepackage{verbatim}
\usepackage[T1]{fontenc}
 \usepackage{aecompl}
\usepackage{color}
\usepackage{multirow}

\usepackage{times}
\usepackage{natbib}

\def\aap{A\&A}

\def\apjl{ApJ}

\def\apjs{ApJS}

\def\bs{\boldsymbol}

\graphicspath{{./fig/}}

\begin{document}

\title[On the possibility of magnetic field detection on a microlensed source star]
{On the possibility of magnetic field detection on a microlensed source star}

\author [Mehrabi et al.]
{Ahmad Mehrabi$^{1}$ \thanks{mehrabi@basu.ac.ir}, Habib khosroshahi$^{2}$ and Hadi Rahmani$^{2,3}$ \\
$^1$ Department of physics, Bu-Ali Sina University, Hamedan, Iran \\
$^2$ School of Astronomy, Institute for Research in Fundamental Sciences (IPM), P.O. Box 19395-5531, Tehran, Iran \\
$^3$ Aix Marseille Universit\'e, CNRS, LAM (Laboratoire d'Astrophysique de Marseille) UMR 7326, 13388, Marseille, France}

\maketitle

\begin{abstract}
In a microlensing event, a large magnification occurs at caustic crossing and provides an opportunity to obtain a stronger signal associated with the object. In this paper we study the possibility of magnetic field detection  in a microlensing event through the Zeeman effect. We follow the prescription introduced by \citep{Robinson:1980} which analyses the spectrum of a star in a Fourier space to deconvolve other broadening mechanism from the actual Zeeman effect. First we study magnification contrast between source and spot in terms of spot size and then consider two distinct strategies using modern spectrographs
and perform a Monte Carlo simulation to find the detection efficiency in magnification-magnetic field plan. The spectral resolution and signal to noise ratio in each strategies, specify suitable places in this plane to detect a magnetic field. 
 Apart from complexity of magnetic field on a star, we consider a simple model and propose a fantastic method to detect magnetic field using spectrum of source stars at caustic crossing.
\end{abstract}


\section{Introduction}
When a compact object (lens) sufficiently aligned with the light source, two images of the source appear due to light bending in the gravitational field. For the galactic events the angular distance of images is of the order of mili-arcsecond,
thus can't  be resolved  using ground based instruments and called micolensing.
This multiplication of images manifests itself as a magnification of the source flux and provides a
powerful tool to study faint objects. For a point like lens the magnification of a point like source is an analytic function but magnification of a uniformly bright circular source is given by elliptical integrals\citep{Witt:1994}. In contrast to a point lens, magnification for a binary system can't be expressed by analytic function and two main approaches have been developed to find the magnification, ray-shooting algorithm \citep{Kayser:1986,Weiss:1987} and adoptive countering \citep{Dominik:2007fv}.

Gravitational microlensing provides a unique tool to study not only the properties of the lens system, 
(e.g extra-solar planets \citep{Cottle:1991zza,Gould:1992aj,Gould:2008zu}), but also the properties of the source star
(e.g stellar atmosphere \citep{Albrow:1998mv,Afonso:2001gh,Gould:2001bg,Abe:2003xb,Fields:2003zx}). Recently author of \citep{Rahvar:2015} has collected the most recent application of microlensing.
For a binary or multiple point like lens systems there are some closed curved in the source plane which the magnification of a point source diverges and called caustic line.
The corresponding images position is called critical curved and formed closed curve like caustic lines. (For more information see \citep{Erld:1993} ).
 Near a caustic line an extended 
source undergoes a large magnification and due to high magnification gradient, the source surface is differentially magnified. 
Relatively slow lens-source proper motion, provides an opportunity to sample the light curve with high cadence and study detail of a source surface. For example the technique has been used to obtain limb-darkening profile of Bulge giant and sub-giant \citep{Fields:2003zx} as well as main sequences \citep{Abe:2003xb}. 

In this paper we study the possibility of measuring the magnetic field which may be existed at the surface of a source star, when the source is crossing the caustic line.
Caustic crossing has two major impacts on the embedded magnetic field, on one hand the spot area will increase and on the other hand total flux is magnified. These two effects increase the signal to noise ratio (SNR) and provide a special time to detect the magnetic field.  In the next paragraph we will shortly discuss the detection of stellar magnetic field.

Evidences for the presence of the Solar magnetic field was established by 
Hale in 1908 \citep{Hale:1908}. He measured the magnetic filed of sunspots to be 
$\sim 2600-2900$ Gauss using the line splitting and polarization.  Detection of the magnetic field over the surface of other stars is much more difficult. There are several phenomenons in the range of magnetic activity that might mimic magnetic field. These effects include dark spots \citep{Berdy:2005}, flares \citep{Favata:1999ja}, enhanced chromospheric emission in some bands \citep{Pizzo:2003,Hall:2004}, and hence, distinguishing magnetic field from these effects is a difficult task. In spite of such difficulty, average magnetic field of Ap stars was 
measured using Zeeman broadening in 1971 \citep{Preston:1971}. \citep{Robinson:1980}  developed an interesting technique to measure the amplitude of the magnetic field using Fourier analysis of the line profile produced due to Zeeman effect. 
His method involved crude assumptions regarding the radiative transfer which later improved by other authors. 
\citep{Saar:1988,Basri:1988,Saar:1990,Ruedi:1997}. 
The detection or measuring the magnetic field has been revolutionised over the past decades due to the developments of the new generation of spectrographs (VLT/UVES, ESO/HARPS, Keck/HIRES, etc.) and spectropolarimeters 
Narval at TBL  and HARPSpol at ESO. 
 More advanced discussions regarding this issue can be found in
 \citep{Mathys:2012} and \citep{Landstreet:2013}.
 To best of our knowledge, there is no attempt to find magnetic field in a microlensing
event. In this work apart from all complexity, we use a simple model for magnetic field on the source star and study the possibility of its detection when the source is crossing a caustic line.

The structure of this paper is as follows. In Sec.(\ref{sec1}) we present the key equation of microlensing and amplification at caustic line. Then we investigate magnification contrast between magnetic and non magnetic area and perform a Monte Carlo simulation to show how much the area of a spot will increase at caustic crossing. In Sec.(\ref{sec2}) we show how a spectral line be modified in the presence of  
a magnetic field and how one can detect the signal of the magnetic field in the spectrum of a star. In Sec.(\ref{sec3}) the possibility of magnetic field detection is studied using Monte Carlo simulation and in Sec.(\ref{con}) we conclude and discuss the results.

\section{Gravitational microlensing and caustic line}\label{sec1}
 Gravitational field of a lens star deflects light of a source star. Such a deflection,
 depending on the mass distribution of lens, may produces two or several images. Since the gravitational field doesn't change the surface brightness of the source,
 magnification is simply the ratio of images to source area. 
 
Using thin sheet approximation for the lens, the deflection angle is given by \citep{sch-1985}: 
\begin{equation}\label{eq-deflection-angle}
\alpha(\bs{x})=\frac{1}{\pi}\int\kappa(\bs{x^\prime})\frac{\bs{x}-
\bs{x^\prime}}{|\bs{x}-\bs{x^\prime}|^2}d^2x^\prime\;,
\end{equation}
where $\bs{x}$ is a 2 dimensional vector normalized to 
\begin{equation}
R_{\rm E}= 
4.0A.U(\frac{M}{M_\odot})^{0.5}(\frac{D_s}{8{\rm Kpc}})^{0.5}(\frac{p(1-p)}{0.25})^{0.5}\;,
\end{equation}
which is the Einstein radius in the lens plane. Where  $D_{s}$ ($D_{d})$ are the source (lens) distance respectively and $p=\frac{D_d}{D_s}$. In Eq.~(\ref{eq-deflection-angle}) the definition of parameters is as following,
\begin{equation}
\kappa(\bs{x})=\frac{\Sigma(R_{E}\bs{x})}{\Sigma_{cr}}~~,~~
\Sigma_{cr}=\frac{c^2}{4\pi G}\frac{1}{p(1-p)D_{s}}\;,
\end{equation} 
where
$\Sigma(\bs{x})$ is the lens surface mass density. 
The lens equation which describe the location of images is given by:
\begin{equation}\label{eq-lens}
\bs{y}=\bs{x} -\alpha(\bs{x})\;,
\end{equation}
where $\bs{y}$ shows a point source in the source plane and normalized to $\frac{D_s}{D_d}R_{E}$.
For a binary system, the lens equation is:
\begin{equation}\label{eq-lens-binary}
\bs{y}=\bs{x} -m_1\frac{\bs{x}-\bs{x}_{1}}{\vert\bs{x}-\bs{x}_{1}\vert^2}
-m_2\frac{\bs{x}-\bs{x}_{2}}{\vert\bs{x}-\bs{x}_{2}\vert^2}\;,
\end{equation}
where $\bs{x}_{1}$ and $\bs{x}_{2}$ are the location of lenses. If the origin of coordinate system set to the center of mass we have
\begin{equation}\label{eq-lens-location}
\bs{x}_{1}=(\frac{dq}{1+q},0)~~,~~\bs{x}_{2}=(\frac{-d}{1+q},0)\;.
\end{equation}
Where $d$ and $q$ are the distance and mass ratio of lenses and without loss of generality we put the lenses along horizontal axes. 
 In addition to $d$ and $q$ parameters,  we need three parameters including 
$u_0$ (minimum impact parameter), $t_0$ (time to reach $u_0$) and $t_E$ (The Einstein crossing time) to completely  describe a binary event.   
The magnification of each image is simply the ratio of image to the source area and is given by
\begin{equation}\label{eq-jacob}
\mu_{i}=det(\frac{\partial\bs{y}}{\partial\bs{x}})^{-1}(\bs{x}_{i})\;, 
\end{equation}
and the total magnification is a sum over individual magnification,
\begin{equation}\label{eq-tot-mag}
\mu(\bs{y})=\sum_{i}det(\frac{\partial\bs{y}}{\partial\bs{x}})^{-1}(\bs{x}_{i})\;.
\end{equation}
Critical points is defined via
\begin{equation}\label{eq-critical}
det(\frac{\partial\bs{y}}{\partial\bs{x}})(\bs{x}_{c})=0\;,
\end{equation}
and form closed curves. On the other hand all points in the source plane $\bs{y}_c=\bs{y}(\bs{x}_c)$, also form closed curves which so called caustic line. When a point source crosses a caustic, according to Eq.(\ref{eq-jacob}), its magnification diverges.
considering finite size effect the magnification doesn't diverge but become very large.  The magnification of an extended source can be 
obtained via inverse rate shooting algorithm\citep{Kayser-1986}. However Dominik introduced an efficient algorithm for calculating the magnification\citep{Dominik:2007fv}.
we use \citep{Dominik:2007fv} algorithm to find the magnification throughout of this paper.
 
Now we consider a simple model to study the magnetic field detection.  We consider a source star with a spot which cover certain amount of projected source area. We parametrize each spot with $f=\frac{A_{s}}{A_{u}}$ which show the initial coverage of spot before crossing the caustic, where $A_{s}$ ($A_{u}$) is the area of spot (source).

When such sources cross the caustic line, 
all parts of the source magnifies but a magnification contrast occur between regions close and far from caustic. Such a magnification contrast increases the signal of the magnetic field when the spot is over caustic so makes it more detectable. We consider a source with a single spot and define the contrast as $\frac{\mu_{\rm s}}{\mu_{\rm u}}$. Where 
$\mu_{\rm s}$ is the magnification of the spotted area and $\mu_{\rm u}$ is the magnification of a uniform source.  
\begin{figure}
\begin{center}
\includegraphics[width=0.5\textwidth]{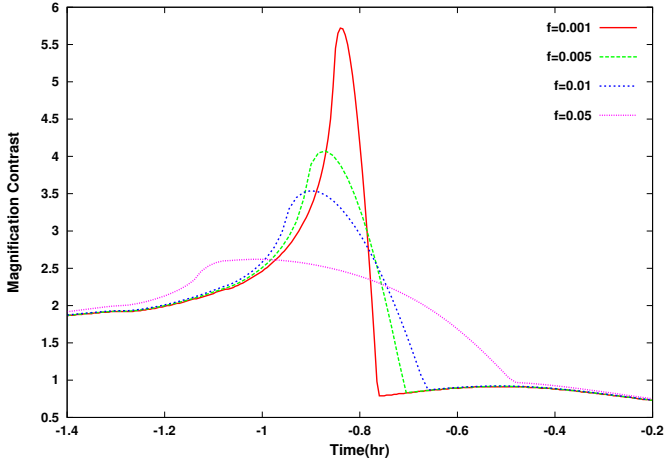}
\caption{Magnification contrast for spots in 4 different sizes in a typical binary lenses event. The red solid, dashed green, tiny dashed blue and dotted pink curves represent magnification contrasts for spot with $f=0.001$, $f=0.005$, $f=0.01$ and $f=0.05$ respectively. }
\label{fig-contrast-caustic}
\end{center}
\end{figure}
In Fig. \ref{fig-contrast-caustic} the magnification contrast is shown for 4 spots with different sizes. As we expected, decreasing the spot size increases the magnification contrast but the initial signal is less than a large spot.
 The amplitude of magnification contrast depends on the location of the spot at caustic crossing, we perform a Monte-Carlo simulation to find probability of occurring a specific magnification contrast for spots with different sizes. In our analyzes, we simulate several events with a spot which located over the source randomly and the magnification contrast is average of this quantity in each event. In this analyze we select a typical binary lenses with ($d=1,~q=0.001,~t_{E}=25~{\rm days}$) and as long as we are near central caustic, our results are not sensitive to change the parameters of lenses. The results of such an exercise for four different $f$ is presented in Fig(\ref{fig-contrast-all}). The contrast can be as large as 10 for 
a spot with initial size $0.001$ while for a large spot high contrasts are less likely.
\begin{figure}
\begin{center}
\includegraphics[width=0.5\textwidth]{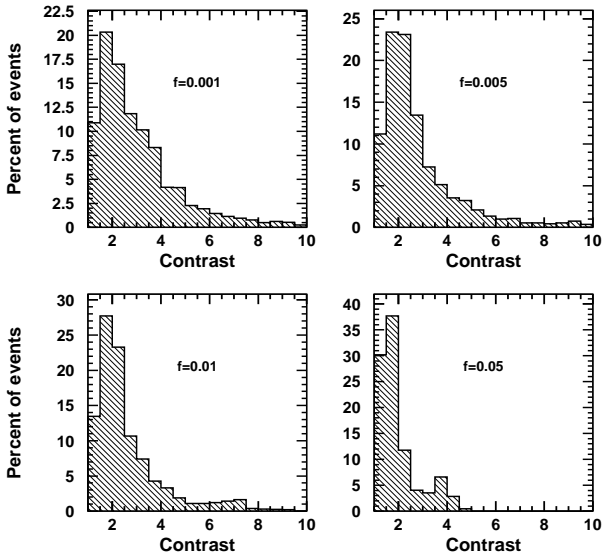}
\caption{Frequency of contrast for spots with different sizes. The mean contrast is calculated as a ratio of spotted to the uniform source magnification around the caustic line (see text for more detail). The contrast for small spot might be larger than a big spot.}
\label{fig-contrast-all}
\end{center}
\end{figure}
In addition to this effect, when a star crosses a caustic line, its brightness might magnify several hundreds times and so the signal will increase significantly. Following, after introducing two strategies, we consider this effect to detect the magnetic field.      

\section{Spectral line distortion in the presence of magnetic field}\label{sec2}
Strong magnetic field could be produced at the surface of stars due to rotation of interior plasma.
These magnetic fields split a sensitive line due to Zeeman effect.
A spectral emission or absorption line is excited if electron makes a transition from 
one state to another. Electron in each energy state has an orbital angular momentum $\vec{L}$ and a spin of $\vec{S}$ resulting in a total angular momentum of $\vec{J}=\vec{L}+\vec{S}$ for the electron.
In the presence of a magnetic field, each state with an angular momentum of $\vec{J}$ splits into $2j+1$ states each with quantum number $M$.
This splitting of energy states is proportional to $Bg$ where  $g$ is the lande factor and depends on the energy levels, orbital and spin angular momentum \citep{reiners-2012}. 
Transition  between two  states  must obey the selection rule $\Delta M=-1,0,1$ and so there are three energy levels at present of magnetic field.
Now we assume $F_0(\lambda)$ be a profile of a sensitive line in the absence of magnetic field, when a magnetic field switch on, it changes to
\begin{equation}\label{eq-mag-pro}
F_{m}(\lambda)=F_0(\lambda)*(\frac{a}{2}\delta(\lambda-\Delta\lambda)+\frac{a}{2}\delta(\lambda+\Delta\lambda)
+ b\delta(\lambda))\;,
\end{equation}
where $*$ indicates convolution and $F_{\rm m}(\lambda)$ and $\delta(x)$ are spectrum profile in the presence of magnetic field and Dirac delta function respectively. In above equation $a$ and $b$ are two constants which depend on direction of magnetic field and line of sight $(\gamma)$, and are given by \citep{Bobcock-1949}:
\begin{equation}\label{eq-a-b-constant}
a = 0.5(1+\cos^2\gamma)~~,~~b = 0.5\sin^2\gamma\;.
\end{equation}
The wavelength displacement due to magnetic field is:
\begin{equation}\label{eq-del-landa}
\Delta\lambda=35.84(\frac{g}{1.2})(\frac{\lambda_0}{0.8\mu m})^2(\frac{B}{1kG}){\rm m\mathring{A}}
\end{equation}
Hare  $\lambda_0$ is the rest frame wavelength of the line. Converting to velocity we have:
\begin{equation}\label{eq-del-vel}
\Delta v=1.34(\frac{g}{1.2})(\frac{\lambda_0}{0.8\mu{\rm m}})(\frac{B}{1kG})~{\rm Km~s^{-1}}
\end{equation}
The typical magnetic field on the surface of stars is on the order of a kilo Gauss which results in velocity shifts of the order of km s$^{-1}$ for $\lambda_0 \sim 0.8\mu{\rm m}$.
According to Eq.~(\ref{eq-del-vel}) infrared and longer wavelength transitions have larger shifts than those of optical. 
For example $1.56\mu m$ of (FeI) and $2.22\mu m$ of (TiI) are two sensitive line which can be used in infrared to detect a magnetic field.   

The brightness profile of a star can be written as:
\begin{equation}\label{eq-profile}
P(\lambda)=F(\lambda)*V(\lambda)*M(\lambda)*I(\lambda)\;,
\end{equation}
where $F(\lambda)$, $V(\lambda)$, $M(\lambda)$ and $I(\lambda)$ indicate the intrinsic profiles, velocity, micro and macro turbulence and the telescope broadening, respectively. In Fourier space Eq.~(\ref{eq-profile}) transforms to
\begin{equation}\label{eq-profile-four}
p(k)=f(k)v(k)m(k)i(k)\;,
\end{equation}
 where convolution converts to ordinary multiplication  in Fourier space.
The intrinsic profile is 
\begin{equation}\label{eq-mag}
F(\lambda)=fF_{m}(\lambda)+(1-f)F_0(\lambda)\;,
\end{equation}
and using Eq.~(\ref{eq-mag-pro}), it transforms to
\begin{equation}\label{eq-mag-four}
f(k)=f_{0}(k)[fa(\cos(2\pi\Delta\lambda k)-1)+1]\;. 
\end{equation}
Our strategy is to measure the magnetic field base on the Robinson method \citep{Robinson:1980}. According to this method, 
by dividing the line profiles of two transitions in Fourier space we have:
\begin{equation}\label{eq-divide-four}
\frac{p(g_1;k)}{p(g_2;k)}=\frac{f(g_1;k)}{f(g_2;k)}\;,
\end{equation}
where all broadening agents from Eq.~(\ref{eq-profile}) will be omitted. Using two lines with small (insensitive) and large (sensitive) lines one can 
easily separate magnetic field signal from other broadening agents. Our following work on the measurement of magnetic field essentially base on the Eq.~(\ref{eq-divide-four}). In this equation $f(g;k)$ comes from Eq.~(\ref{eq-mag-four}) and depend on the $f_{0}(k)$ and $\Delta\lambda$ where $f_{0}(k)$ is the Fourier profile for $B=0$ and $\Delta\lambda$ as a function of $g$ that is given by Eq.~(\ref{eq-del-landa}). 
 To find $f(g;k)$ we need to calculate Fourier transform for a discrete sample. According to Nyquist theorem if sampling rate in velocity space be $\Delta v$, its Fourier transform is confined
in the range$[-\frac{1}{2\Delta v},\frac{1}{2\Delta v}]$. This limitation imply that we can't detect magnetic field with
$\Delta\lambda/\lambda <<\Delta v/c$, as we expected.
 
\section{Detectability of the magnetic field at caustic crossing}\label{sec3}
In this section we study the detectability of the magnetic field at caustic crossing. 
To do so we consider two sets of spectroscopic observations as following: 
\begin{enumerate}
 \item spectral resolution of $\Delta v=1$ {Km} {s}$^{-1}$ with a signal to noise ration (SNR) of 20
  \item spectral resolution of $\Delta v=5$ {Km} {s}$^{-1}$ 
  with  SNR = 50
\end{enumerate}
\noindent and perform a Monte Carlo simulation to study the detectability of magnetic field at caustic crossing. These are a typical value for a spectrogragh and in this paper we aim to study the effects of caustic crossing on field detection.  We simulate several 
events with different parameters $(q,d,u_0,t_E)$ and near central caustic (high magnification), the detectability of a magnetic field is calculated. Moreover in this study we assume that the spot can locate over each part of the source with equal probability so the location of spot is chose randomly in each event. Now we need to define a criteria to select detectable events. Since the signal of magnetic field comes from Eq.~(\ref{eq-mag-four}), by Assuming 
a sensitive line with $(\lambda_{0}=0.8 ~\mu m,~g_1=2.5)$  and an insensitive line with $(\lambda_{0}=0.8~\mu m,~g_2=1)$, we have:
\begin{equation}\label{eq:signal}
\frac{f(g_1;k)}{f(g_2;k)}=\frac{f_{0}^{sen}(k)}{f_{0}^{ins}(k)}\frac{[fa(\cos(2\pi\Delta\lambda_{sen} k)-1)+1]}{[fa(\cos(2\pi\Delta\lambda_{ins} k)-1)+1]}\;.
\end{equation}
 where superscript $sen$ and $ins$ stand for sensitive and insensitive lines,  respectively. In the following we assume that $f_{0}^{sen}(k)$ and $f_{0}^{ins}(k)$ are known function that in principal can be obtained by solving the radiative equation of transfer \citep{Saar:1988}. In this case the signal comes from function
\begin{equation}\label{eq:signal2}
S(B)=\frac{fa(\cos(2\pi\Delta\lambda_{sen} k)-1)+1}{fa(\cos(2\pi\Delta\lambda_{ins} k)-1)+1}\;.
\end{equation} 
The value of $a$ is given by Eq.~(\ref{eq-a-b-constant}) and we assume an average value $\gamma=34$ from \citep{Marcy:1981}. 
For $B=0$, we have $S(B=0)=1$, so any deviation from 1 indicates $B\neq 0$. 
  We assume a Gaussian distribution for flux errors which is distributed like true space in Fourier space, then among all simulated events, those with signal more than $5\sigma$ are flagged as detectable events. We should notice that in our analysis $f$ is a function of time and we use a mean value for this quantity (see Fig. \ref{fig-contrast-caustic}). This mean value is obtained by averaging $f(t)$ around caustic line.
\begin{figure}
\begin{tabular}{c }
\includegraphics[width=0.45\textwidth]{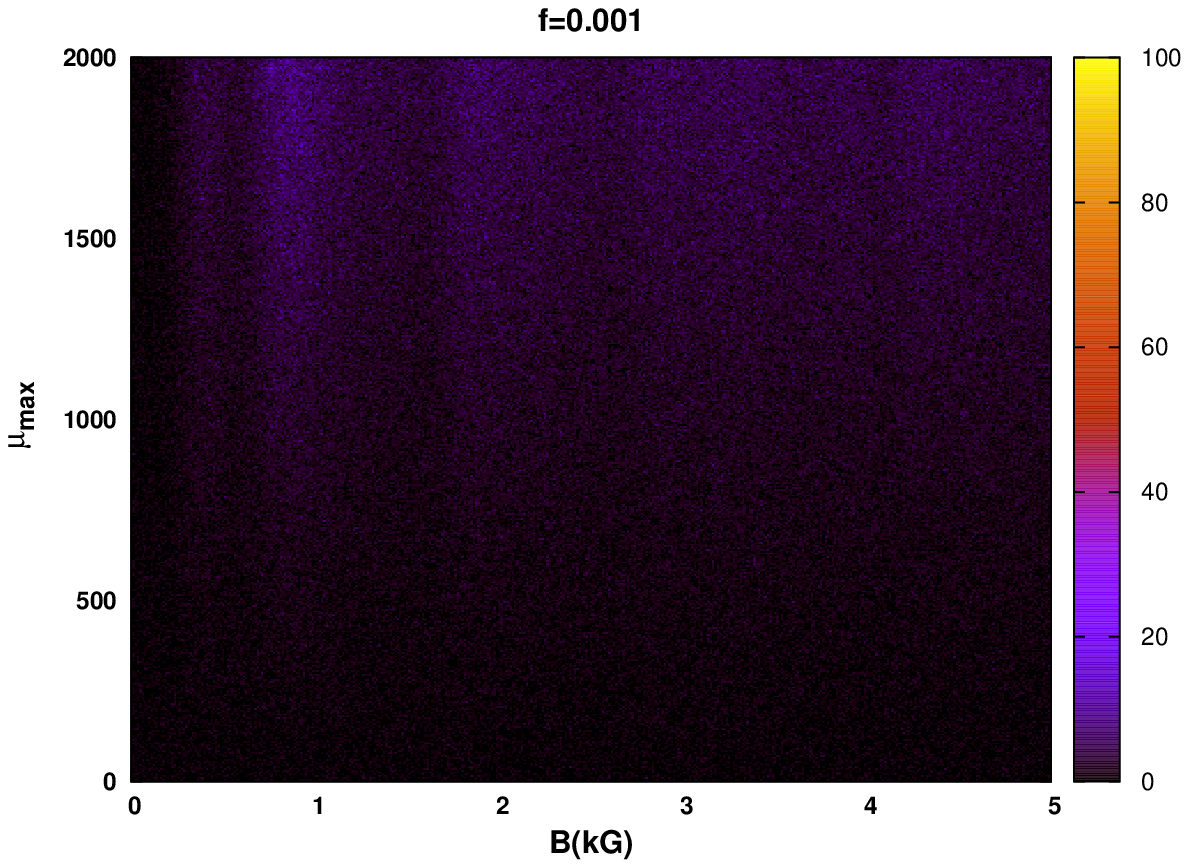} \\
\includegraphics[width=0.45\textwidth]{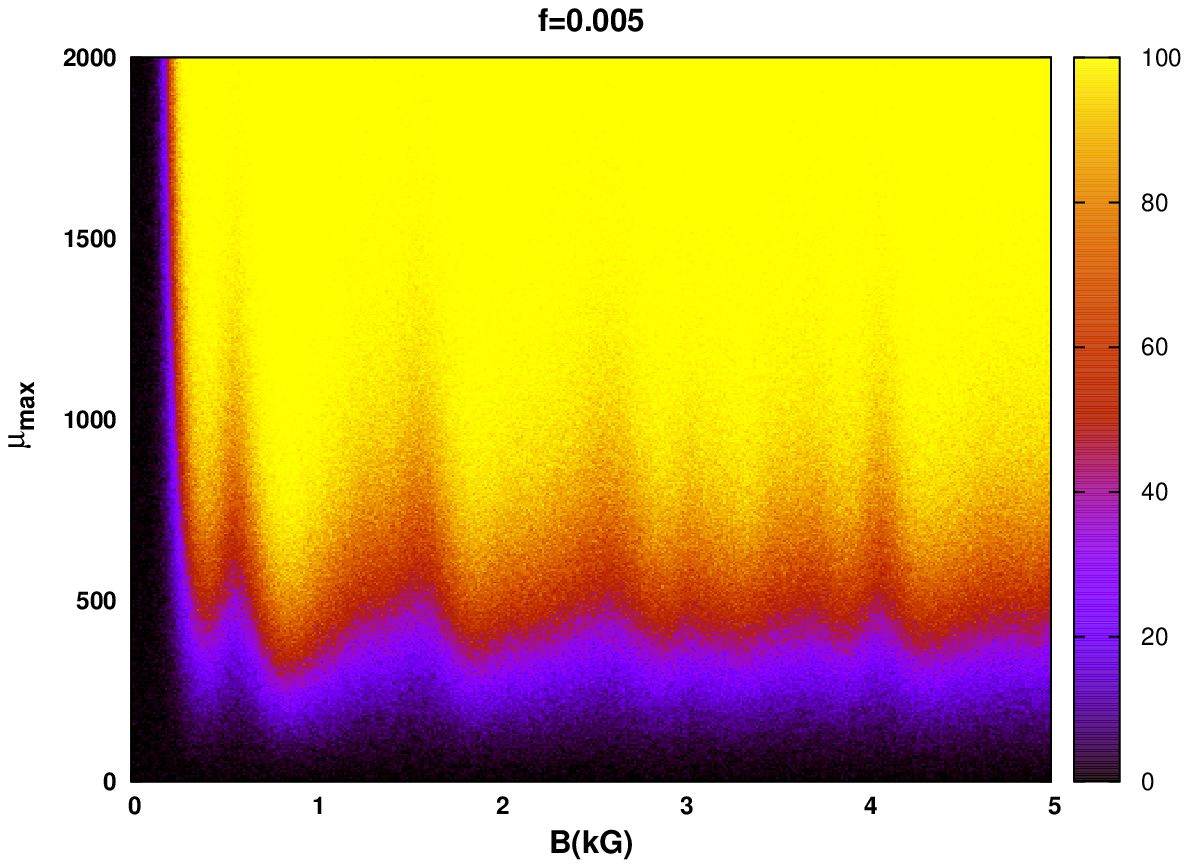}\\
\includegraphics[width=0.45\textwidth]{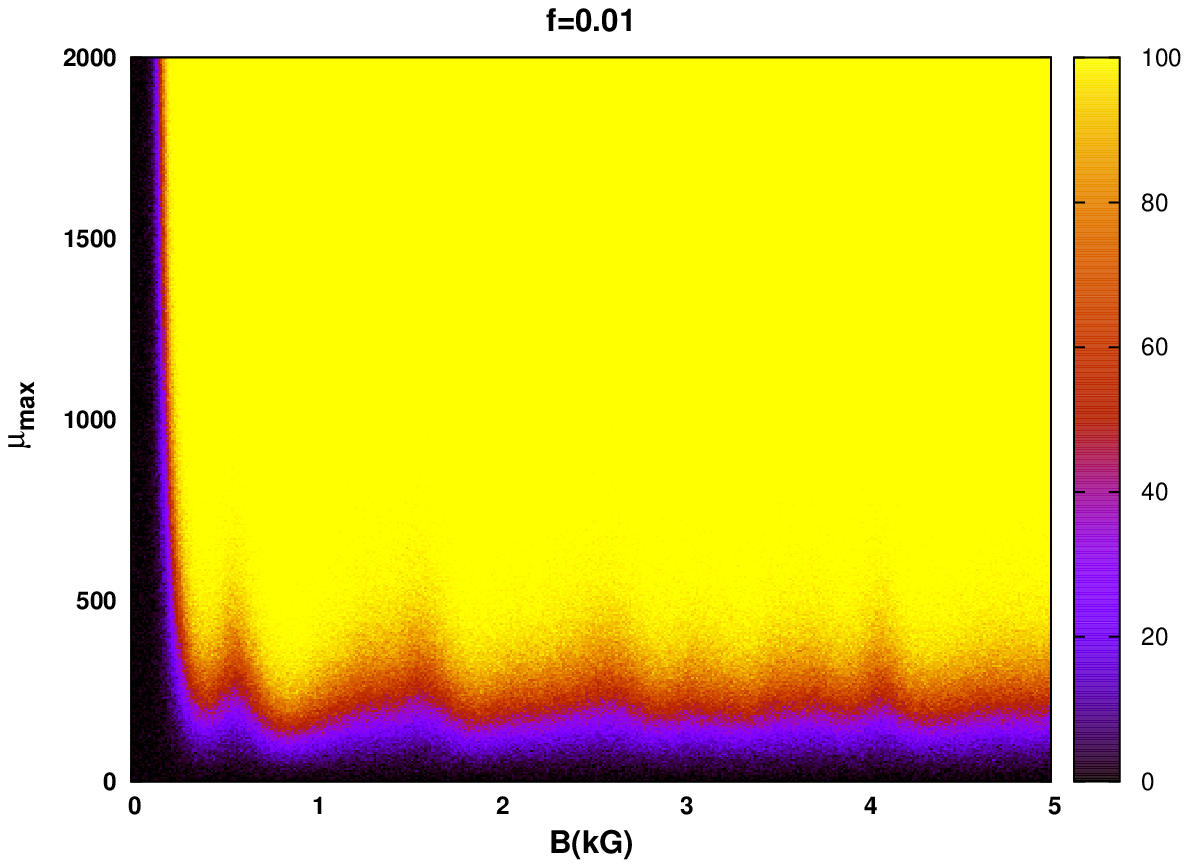}\\
\includegraphics[width=0.45\textwidth]{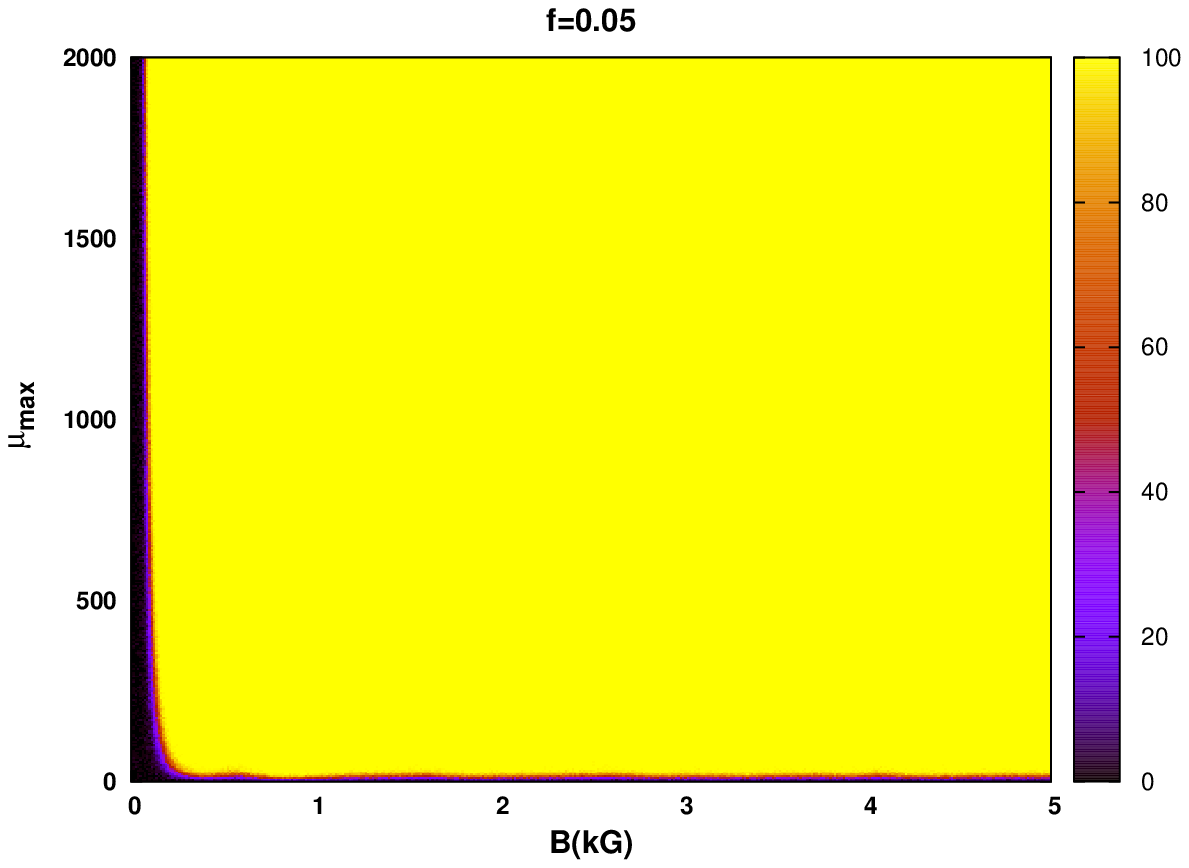}\\
\end{tabular}
\caption{Percent of detectable event in the $A_{\rm max}-B$ plane. Dark area (yellow point) present undetectable (fully detectable) events. The size of spots is presented above each diagram.} 
\end{figure}\label{fig-stt1}

The magnetic field detectability is defined as a number of events which pass our criteria per total number of events and the results for strategy 1 and 2 are shown in Figs.~(3)  and (\ref{fig-stt2}). In strategy 1, for $f=0.001$ the detectable events are very rare and for very high magnification the detectability is less than $20\%$. For a spot with initial size $f=0.005$ magnification larger than around 500 is required to detect magnetic field but for $f=0.05$ events with small magnification are fully detectable due to high signal from a large spot. The results for strategy 2 are almost like the results of strategy 1 but due to low spectral resolution, small magnetic field can't be detected while high SNR help to detect magnetic field for $f=0.001$ in contrast with strategy 1. 

\begin{figure}
\begin{tabular}{c }
\includegraphics[width=0.45\textwidth]{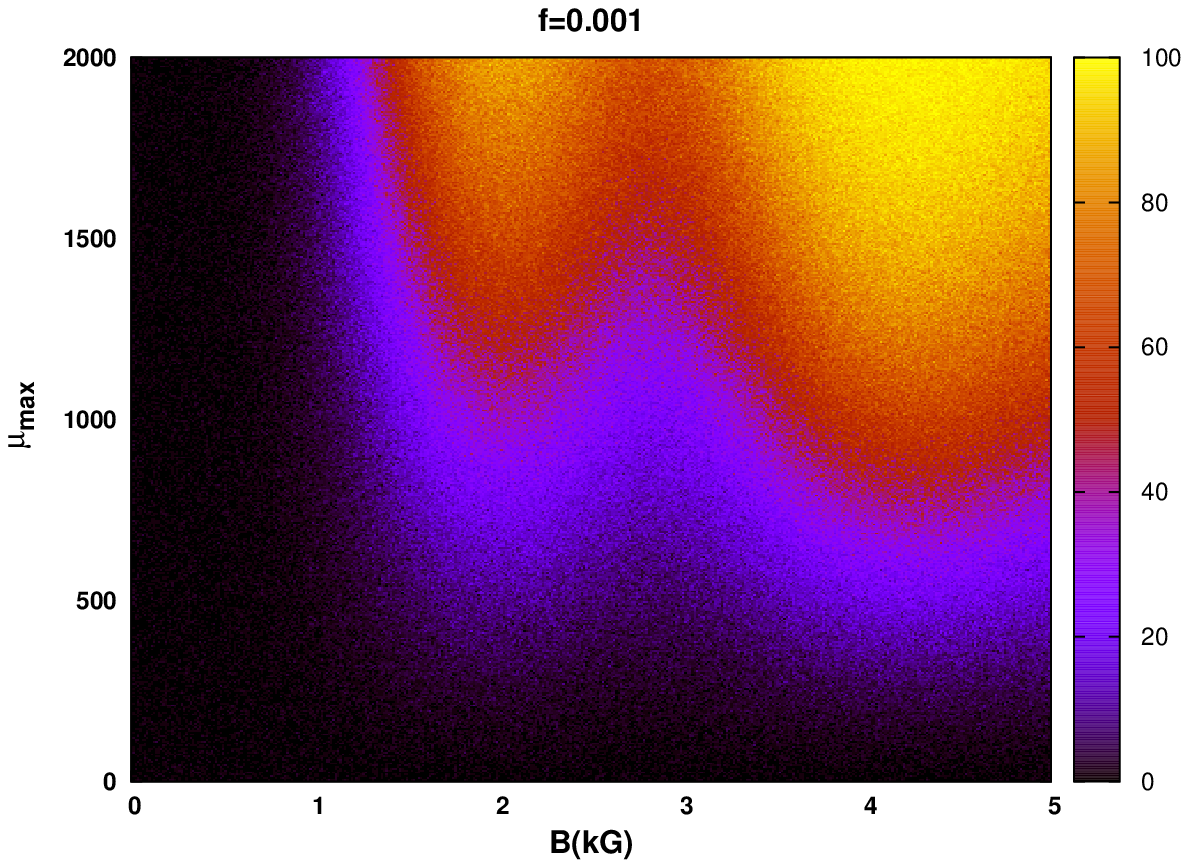} \\
\includegraphics[width=0.45\textwidth]{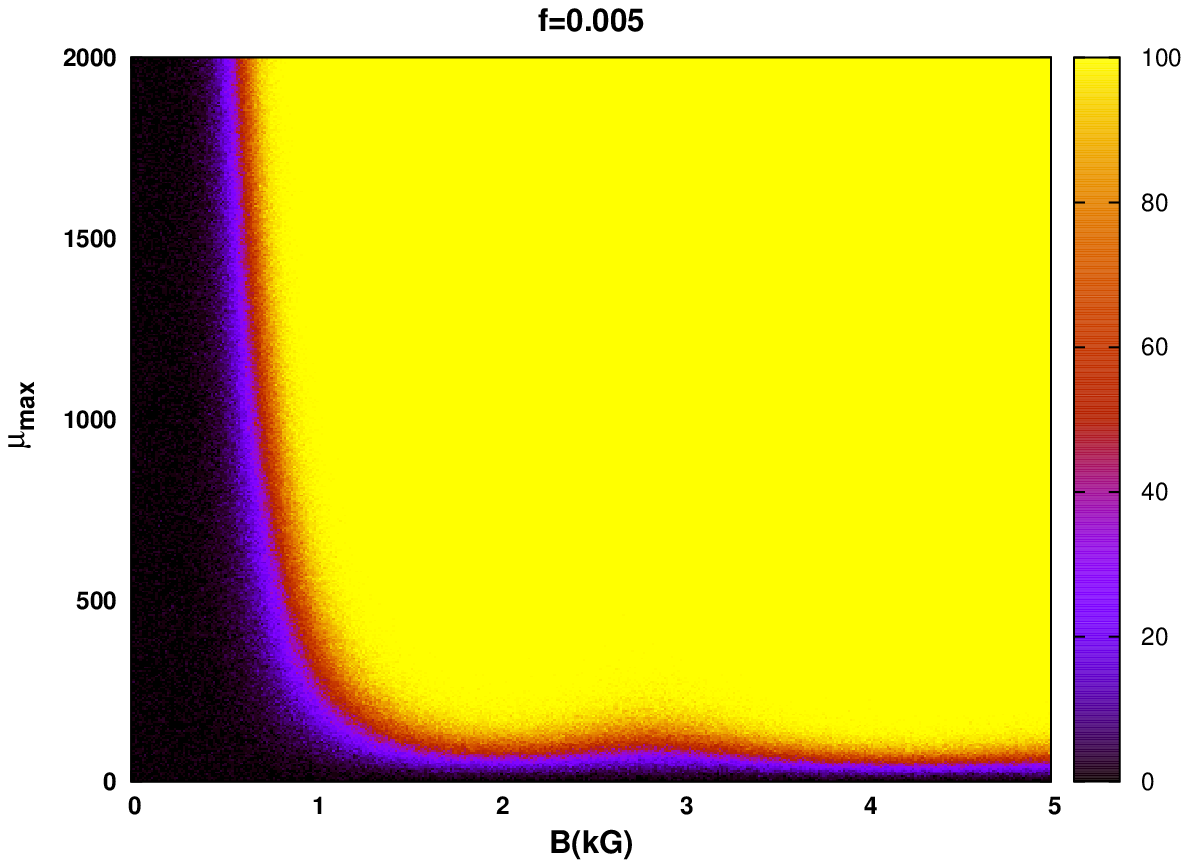}\\
\includegraphics[width=0.45\textwidth]{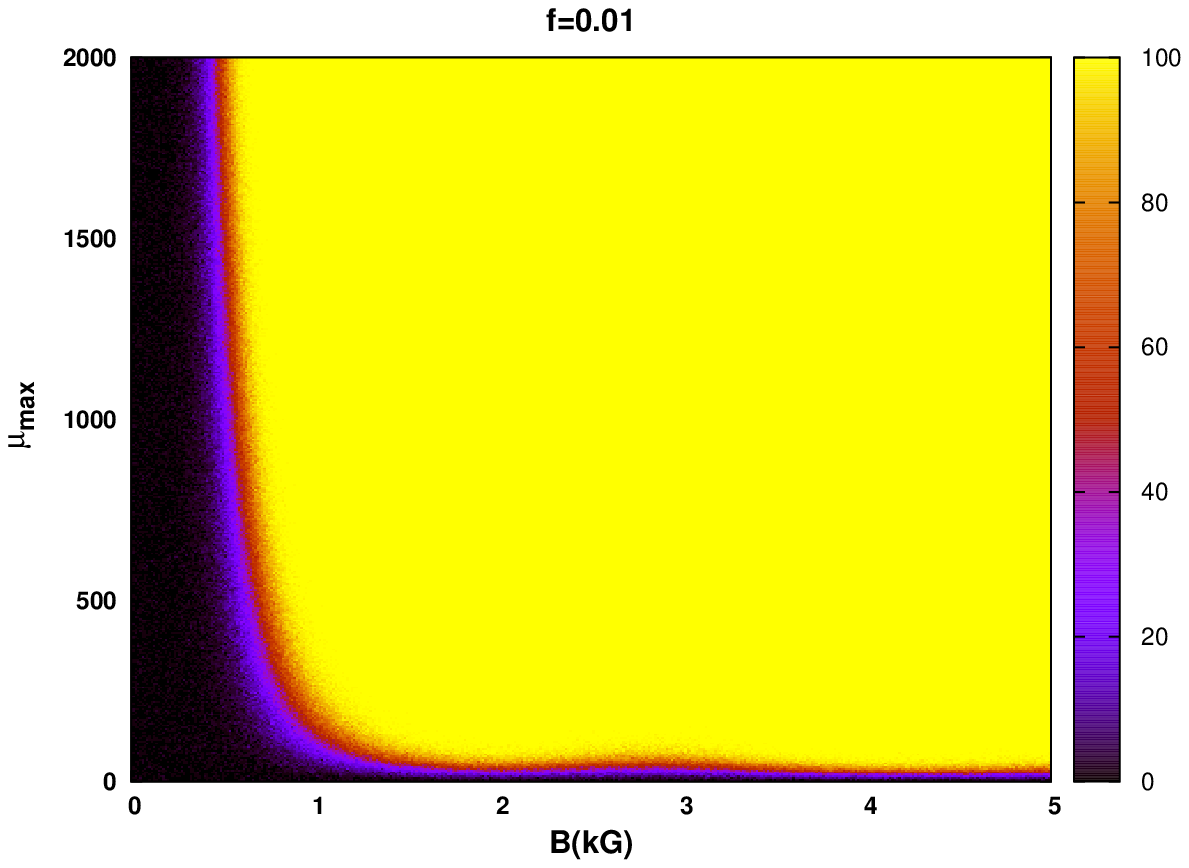}\\
\includegraphics[width=0.45\textwidth]{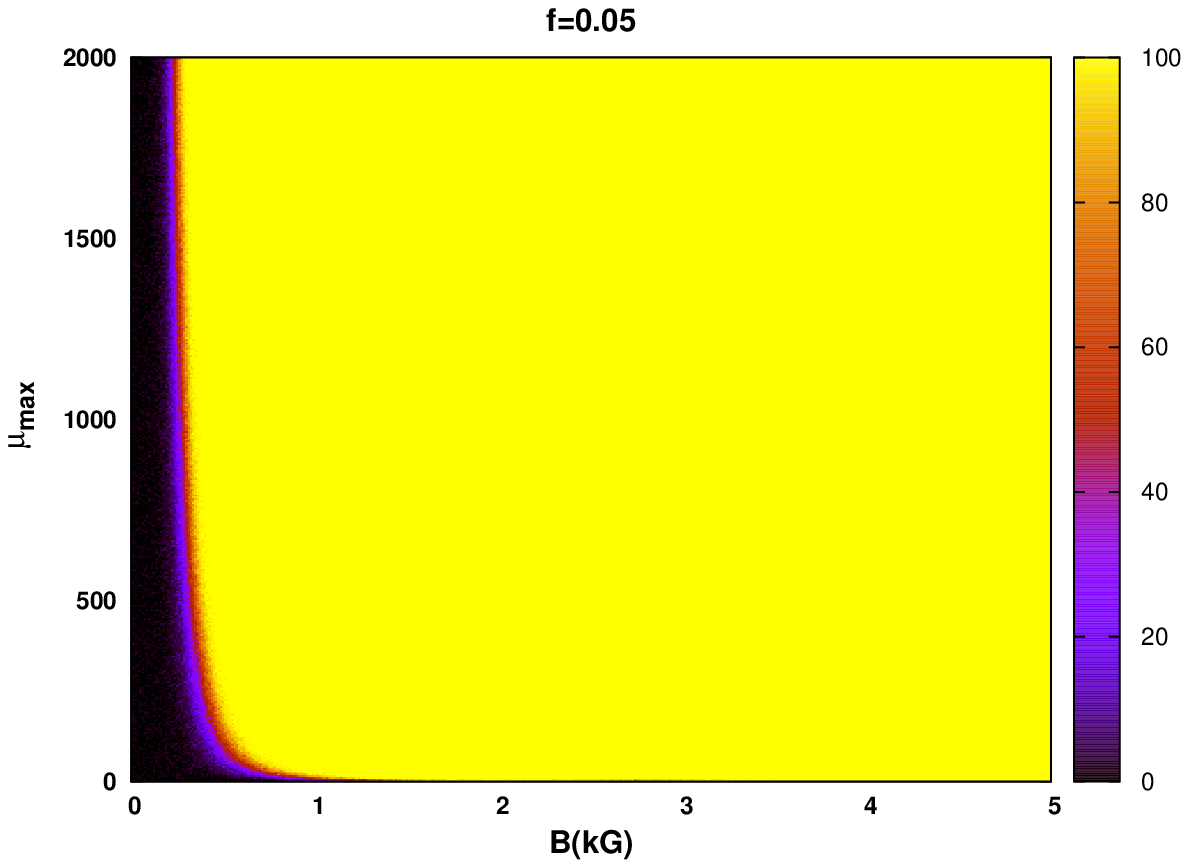}\\
\end{tabular}
\caption{Same as Fig.~(3) but for strategy 2.} 
\end{figure}\label{fig-stt2}

 Since in a microlensing event it is possible to realize type of a  source star before caustic crossing, selecting a more sensitive star leads to a higher chance to detect magnetic field.  M-type stars have radii smaller than solar radius and also rotate slower so these conditions lead to  produce  a large field in such stars. 
In addition two unwanted  broadening mechanism, rotational and temperature broadening, in a M-type star are smaller compare to a sun like star. All of these effects will facilitate field detection in an M-type source star. (for more information on the magnetic field of M-type star see \citep{Reiners:2006dn,Reiners:2010}). Sun like stars may be active or less active but it is not possible to find out that a sun like source is active before caustic crossing. (to see a list of field detection on a sun like star refer to \citep{reiners-2012} ). Moreover giants stars can be rather active. The field of giants are much weaker than dwarfs but the size of spots may be very large \citep{Petit:2004pq}. Giants stars also are good candidate to detect magnetic field.

\section{Conclusion}\label{con}
Several teams including OGLE, EROS, MOA, PLANET and MicroFUN are observing sky to find a microlensing events. These surveys detect several thousand events per year and a fraction of them are binary system. To our knowledge there was no attempt to study the magnetic field of source star in these events. In this work we propose an interesting approach to detect magnetic field which may be existed on the surface of the source. Our detection mechanism based on the Robinson method which investigate two line profiles of a star in Fourier space to deconvolve all broadening agents from magnetic one. We study the possibility of magnetic field detection when a source is crossing the caustic line.
At a caustic crossing not only the flux of a source increase which results higher SNR, but also the coverage fraction of spot 
increases due to magnification gradient. We consider two spectroscopic strategies which are in agreement with recent powerful spectrograph, and find the detectability of magnetic field in the range $B\in(0-5){\rm kG}$ and magnification in the range $\mu_{\rm max}\in(10-2000)$. In addition, in a microlensing events, is is possible to find out the type of source stars before caustic crossing. This situation provide possibility of a more active source selection. Among all stars, M-type stars and giants are two good candidate to detect a magnetic field. 
Our results show that high magnification channel provide an unique opportunity to study the magnetic field of stars which are located at bulge and disk.  

\section*{Acknowledgments}
We should thank 	Sohrab Rahvar for his useful comments. Also we notice that the basic idea of magnetic field detection in a microlensing event is proposed by Sohrab Rahvar.
 
{\footnotesize
\bibliographystyle{mnras}
\bibliography{ref}
}

\bsp

\label{lastpage}

\end{document}